\documentclass[twocolumn]{aastex62}

\received{}
\revised{}
\accepted{}
\submitjournal{ApJ}

\shorttitle{Eccentric BBH Population}
\shortauthors{Fang et al.}

\usepackage{amsmath, amsthm, amssymb,slashed,bm}
\usepackage{graphicx}
\usepackage{epstopdf}
\usepackage{multirow}
\usepackage{subfigure}

\usepackage[utf8]{inputenc}

\newcommand{\eg}{e.g.}
\newcommand{\ie}{i.e.}

\begin{document}


\title{The Population of Eccentric Binary Black Holes: Implications for mHz Gravitational Wave Experiments}

\correspondingauthor{Xiao Fang}
\email{xfang@email.arizona.edu}

\author[0000-0002-5054-9566]{Xiao Fang}
\affil{Department of Astronomy and Steward Observatory, University of Arizona, Tucson, AZ 85719, USA}
\affil{Department of Physics, The Ohio State University, Columbus, Ohio 43210, USA}
\affil{Center for Cosmology and AstroParticle Physics, Department of Physics, The Ohio State University, Columbus, Ohio 43210, USA}

\author{Todd A. Thompson}
\affil{Department of Astronomy, The Ohio State University, Columbus, Ohio 43210, USA}
\affil{Center for Cosmology and AstroParticle Physics, Department of Physics, The Ohio State University, Columbus, Ohio 43210, USA}
\affil{Institute for Advanced Study, Princeton, New Jersey 08540, USA}

\author{Christopher M. Hirata}
\affil{Department of Physics, The Ohio State University, Columbus, Ohio 43210, USA}
\affil{Department of Astronomy, The Ohio State University, Columbus, Ohio 43210, USA}
\affil{Center for Cosmology and AstroParticle Physics, Department of Physics, The Ohio State University, Columbus, Ohio 43210, USA}




\begin{abstract}
The observed binary black hole (BBH) mergers indicate a large Galactic progenitor population continuously evolving from large orbital separations and low gravitational wave (GW) frequencies to the final merger phase. We investigate the equilibrium distribution of binary black holes in the Galaxy. Given the observed BBH merger rate, we contrast the expected number of systems radiating in the low-frequency $0.1-10$\,mHz GW band under two assumptions: (1) that all merging systems originate from near-circular orbits, as may be indicative of isolated binary evolution, and (2) that all merging systems originate at very high eccentricity, as predicted by models of dynamically-formed BBHs and triple and quadruple systems undergoing Lidov-Kozai eccentricity oscillations. We show that the equilibrium number of systems expected at every frequency is higher in the eccentric case (2) than in the circular case (1) by a factor of $\simeq2-15$. This follows from the fact that eccentric systems spend more time than circular systems radiating in the low-frequency GW bands. The GW emission comes in pulses at periastron separated by the orbital period, which may be days to years. For a LISA-like sensitivity curve, we show that if eccentric systems contribute significantly to the observed merger rate, then $\simeq10$ eccentric systems should be seen in the Galaxy. 
 \end{abstract}


\section{Introduction}
The gravitational wave (GW) experiments LIGO and VIRGO \citep{2009RPPh...72g6901A,2012JInst...7.3012A} have recently made the first discoveries of binary black hole (BH) and neutron star (NS) mergers, and they are expected to detect many more such systems in future observing runs \citep{2016PhRvL.116f1102A,2016PhRvL.116x1103A,2017PhRvL.118v1101A,2017ApJ...851L..35A,2017PhRvL.119n1101A,2017PhRvL.119p1101A,2018arXiv181112907T}. Meanwhile, upcoming GW experiments, \eg, LISA \citep{2017arXiv170200786A}, DECIGO \citep{2006CQGra..23S.125K}, Taiji \citep{2015JPhCS.610a2011G}, and TianQin \citep{2016CQGra..33c5010L} will focus on lower frequency ranges, and thus different types and evolutionary stages of compact object binaries.

Current high-frequency GW searches focus on binaries with circular orbits, which are expected both from isolated massive star binary evolution models and the circularization of the orbit during GW inspiral. Although the condition that the orbital eccentricity is small in the LIGO band may well be satisfied, it may not be true for systems at much lower frequency, where formation channels different from isolated binary evolution may imprint themselves \citep[\eg,][]{2016PhRvL.116w1102S,2017ApJ...842L...2C}. In particular, two alternative channels for the production of merging compact object binaries have been suggested, and both predict large eccentricities ($e\gtrsim 0.9$) when the system is radiating GWs at low frequency ($\lesssim 1\,$mHz): (1) dynamically formed compact object binaries within and ejected from globular clusters and other dense stellar systems \citep[\eg,][]{2016PhRvD..93h4029R,2017ApJ...834...68C,2017ApJ...836L..26C,2017MNRAS.467..524B,2018MNRAS.481.5123B,2018MNRAS.473..909B,Samsing_DOrazio,DOrazio_Samsing,2018arXiv181104926R,2018PhRvL.120s1103K,2018arXiv181111812K,2018ApJ...860....5G,2018arXiv180506458A,2018PhRvL.121p1103F,2018arXiv181103640A} and (2) hierarchical triple and quadruple systems undergoing Lidov-Kozai (LK) eccentricity oscillations \citep[\eg,][]{2002MNRAS.330..232C,2003ApJ...598..419W,2002ApJ...578..775B,2011ApJ...741...82T,2012ApJ...757...27A,2013ApJ...773..187N,2017arXiv170609896H,2016ApJ...816...65A,2016ApJ...828...77V,2017ApJ...841...77A,2017ApJ...836...39S,2017ApJ...846..146P,2018MNRAS.476.4234F,2018MNRAS.478..620H,2018ApJ...864..134R,2018arXiv180508212R,2018arXiv181110627F,2018ApJ...863...68L}. Note that the actual frequency range of the ``highly-eccentric'' systems depends on the formation mechanism, and some mechanism can produce highly-eccentric systems at even higher frequencies ($\gtrsim 10\,$mHz), \eg, the highly eccentric GW capture channel in clusters and the evection-induced migration in hierarchical systems.

Here, we argue that if the eccentric channels proposed account for the observed BBH merger rate, then there must be a large population of highly-eccentric BBHs waiting to be discovered at low GW frequencies, in the $0.1-10$\,mHz band. In direct analogy with \cite{2012ApJ...750..106S}, who studied the population of tidally-interacting high-eccentricity migrating hot Jupiters, in a steady state, the observed merger rate together with the continuity equation directly yields the number of eccentric BBH systems in the Galaxy. Since the periastron distance is directly related to the frequency of maximum GW power, these systems will appear as short pulses spaced in time by orbital period. Because the binary orbital angular momentum is approximately conserved at high eccentricity, the periastron distance and peak GW frequency are also nearly invariant during the high-eccentricity ``migration" from large to small semi-major axis.

Here, we consider the possibility that a fraction of the observed LIGO events arise from a highly eccentric initial state. Assuming that the birth and death rates of BBHs are in equilibrium, the continuity equation yields the distributions of orbital properties at every GW frequency, as BBHs evolve toward coalescence. We use the equilibrium assumption to derive the distributions of orbital elements for circular and highly-eccentric systems. Because highly eccentric binaries spend more time radiating at a given GW frequency than a circular system with the same masses (\S \ref{sec:circ_ecce}), the equilibrium number of systems in the Galaxy is larger for eccentric systems than for circular systems if both channels contribute equally to the observed LIGO merger rate. Depending on the initial period (semi-major axis) distribution assumed for the eccentric channel, the equilibrium ratio of eccentric systems to circular systems in the $\sim 0.1-10$\,mHz band is $\sim2-15$. These eccentric systems have peak GW frequencies from $0.1-10$\,mHz, with orbital periods of order days or months, and Galactic systems can be detected by future GW interferometers.

 In \S\ref{sec:circ_ecce}, we show analytically that the equilibrium number of eccentric systems should outnumber the equilibrium number of circular systems under generic assumptions. In \S \ref{sec:results}, we calculate the distribution of BBHs as a function of GW frequency for several progenitor populations, including dynamically-formed BBHs in dense stellar clusters, and triple systems undergoing LK oscillations. In all cases, we find an enhancement in the number of eccentric systems relative to circular systems in the $0.1-10$\,mHz band. In \S\ref{sec:discussion}, we discuss the astrophysical implications of the possible existence of a large population of eccentric BBHs, and their detectability.

\section{Circular Versus Eccentric BBH Populations}
\label{sec:circ_ecce}

Assuming that the birth and death rates of BBHs are in equilibrium, the continuity equation yields the distributions of orbital properties at every GW frequency, as BBHs evolve toward coalescence. Under this equilibrium assumption, the distribution of any quantity $x$ is simply given by the chain rule
\begin{equation}
\frac{dN}{dx} = \frac{dN}{dt}\frac{dt}{dx} = \frac{\Gamma}{\dot{x}}~,
\end{equation}
where we have defined the steady ``inflow'' and ``outflow'' rate of systems as $\Gamma$.

For BBHs, the variations in the orbital parameters as the systems evolve  depend sensitively on the eccentricity. The time-averaged evolution of the semi-major axis $a$ and eccentricity $e$ for a binary of masses $m_1$ and $m_2$ due to GW emission are \citep{1964PhRv..136.1224P}
\begin{align}
\left\langle\dot{a}\right\rangle &= -\frac{64}{5}\frac{G^3m_1m_2M}{c^5a^3(1-e^2)^{7/2}}\left(1+\frac{73}{24}e^2+\frac{37}{96}e^4\right)~,\label{eq:dadt} \\
\left\langle\dot{e}\right\rangle &= -\frac{304}{15}e\frac{G^3m_1m_2M}{c^5a^4(1-e^2)^{5/2}}\left(1+\frac{121}{304}e^2\right)~,\label{eq:dedt}
\end{align}
where $M=m_1+m_2$, $G$ is the Newton's constant, and $c$ is the speed of light.

\begin{figure*}
\centering
\includegraphics[width=9cm]{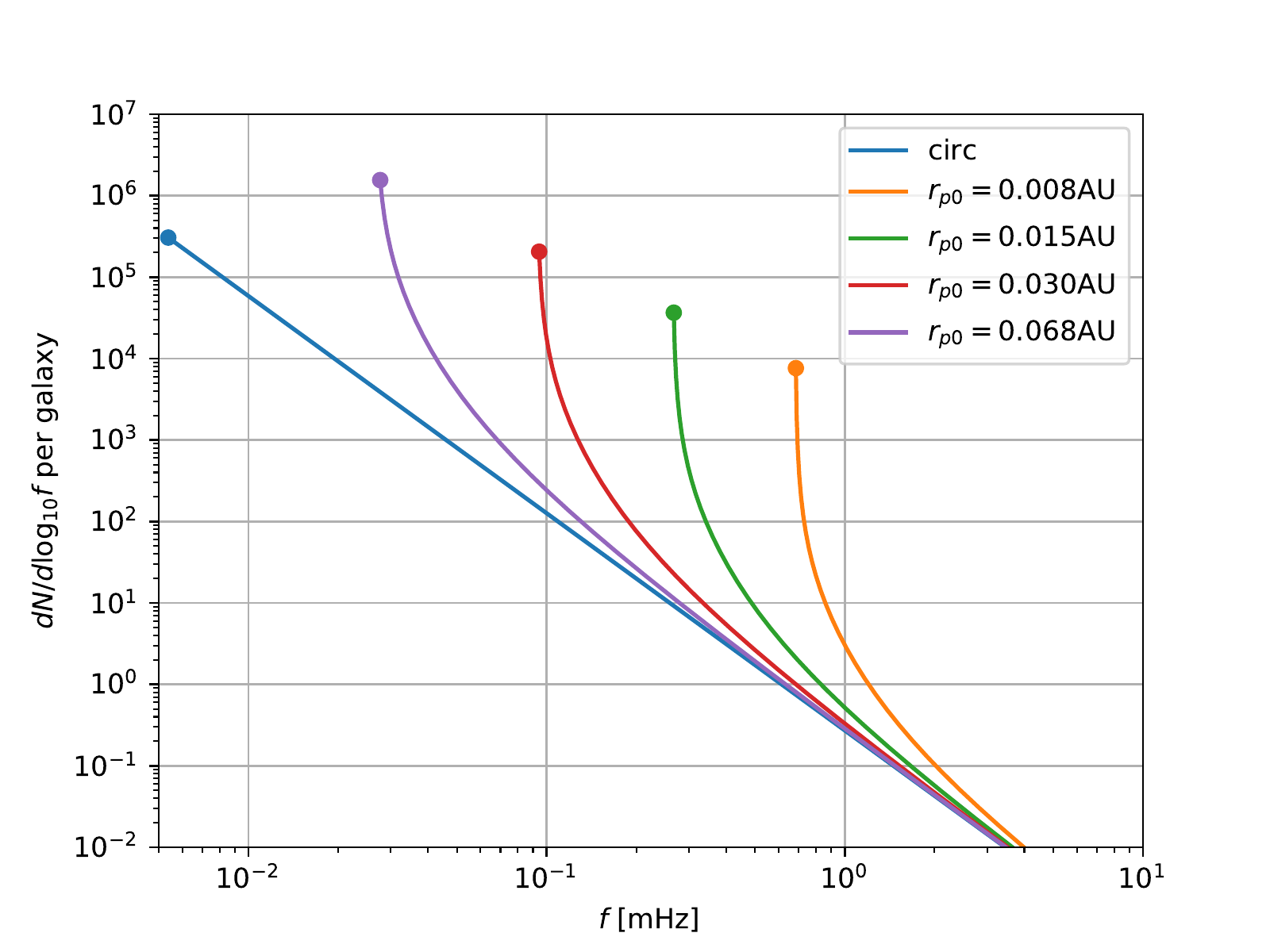}\includegraphics[width=9cm]{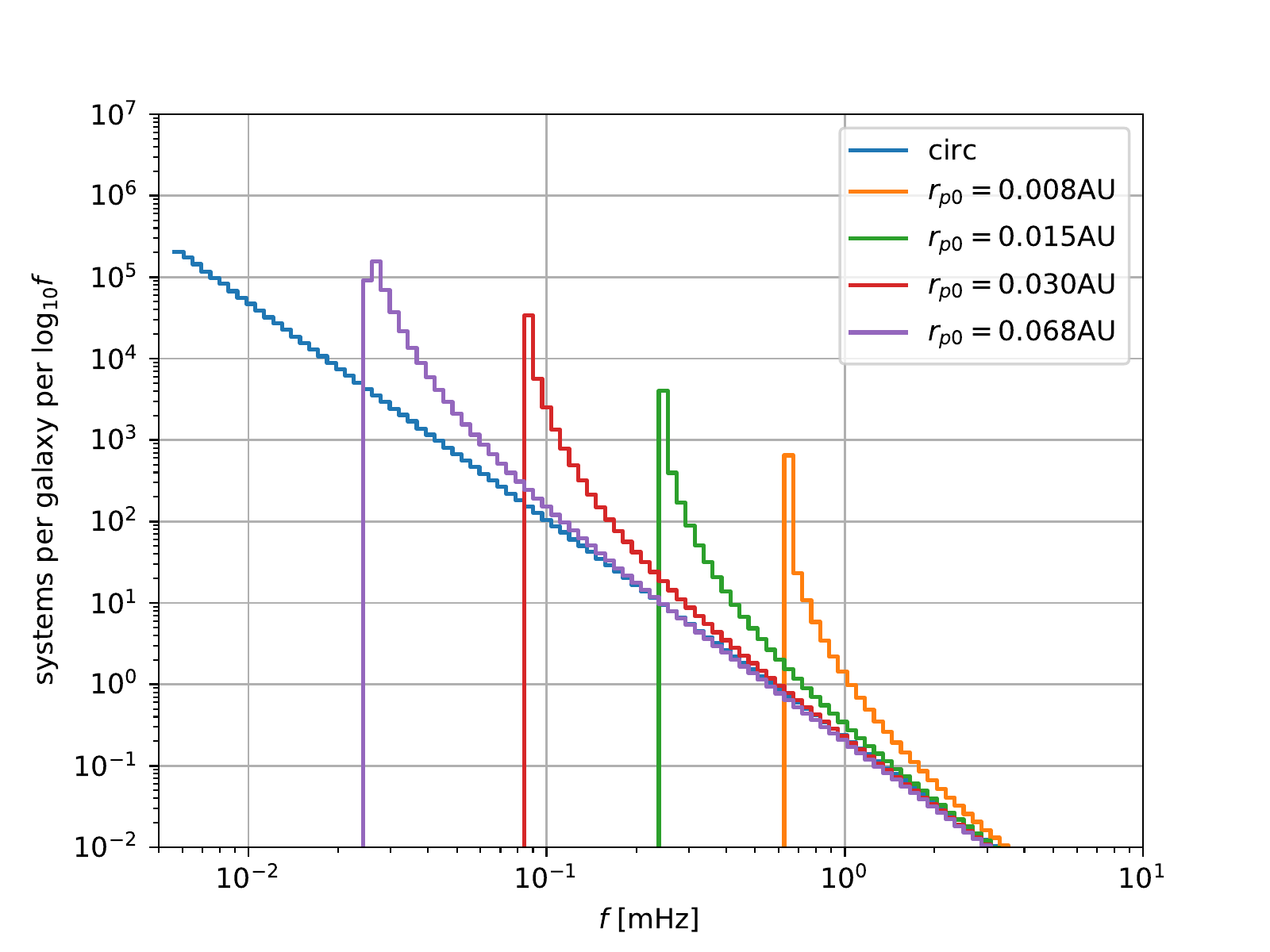}
\caption{{\it Left:} Frequency distributions ($dN/d\log_{10}f$) per Milky Way-size galaxy, for cases where eccentric and circular merger rates are the same as the observed LIGO rate and the initial configurations for the eccentric populations are fixed at $a_0=1\,$au, and $r_{p0}=0.008$, $0.015$, $0.030$, $0.068$\,au, corresponding to merger times of $\simeq 6\times 10^6$, $5\times 10^7$, $6\times 10^8$, and $10^{10}$\,yr, respectively. The dots mark the initial values. For comparison, the circular case is shown in the blue line, which starts at $t_{\rm gw}=10^{10}$yr. {\it Right:} The corresponding peak frequency histograms normalized to per $\log_{10}f$ bin. The number of systems in each frequency bin is an integration of the distribution over that bin, which yields much smaller enhancement than the ratios of distributions as shown on the left and strongly depends on the bin spacing. Also note that the huge enhancement ratios between eccentric and circular cases are highly peaked at initial frequencies and will be smeared out due to the broad distribution of realistic systems.}
\label{fig:dndlogf_singles_logscale2}
\end{figure*}

For the circular binary case ($e=0$), the frequency of the GWs, $f$, is set by the orbital period $P$, hence the semi-major axis $a$, \ie, $f=2/P=2 \sqrt{GM/(4\pi^2 a^3)}~$. Combining $f$ with Eq.~(\ref{eq:dadt}), we obtain the time-averaged rate of increase in the GW frequency $\dot{f}/f = -3\dot{a}/(2a) = 96G^3m_1m_2M/(5c^5a^4)$. If the LIGO BH mergers are produced by initially circular binaries, there exists a distribution of circular BBHs at each $a$ and $f$. If the distribution is in equilibrium, then the frequency distribution, $dN/df$, should be proportional to $dt/df=1/\dot{f}$, and it is normalized by the merger rate $\Gamma$, \ie, \citep[as in \eg,][]{2003MNRAS.346.1197F}
\begin{align}
\left.\frac{dN}{df}\right\vert_{\rm circ} = \frac{\Gamma}{\dot{f}} = \frac{5\Gamma}{96}\frac{c^5}{G^3m_1m_2M}\left(\frac{4GM}{4\pi^2}\right)^{4/3} f^{-11/3}~.
\end{align}

For the eccentric binary case ($e>0$), the GW frequency varies with period $P$, and the peak frequency $f_p$ is set by the periastron distance $r_p=a(1-e)$, \ie, $f_p=2 \sqrt{GM/(4\pi^2 r_p^3)}~$. The rate of increase in $f_p$ is then $\dot{f}_p/f_p = -3\dot{r}_p/(2r_p)$. In the highly eccentric limit ($e\rightarrow 1$), we have
\begin{align}
\dot{r}_p &= -\frac{59}{3}\frac{1}{2^{7/2}}\frac{G^3m_1m_2M}{c^5}\frac{1}{(a r_p)^{3/2}}~,\\
\frac{\dot{f}_p}{f_p} &= \frac{59}{2^{9/2}}\frac{G^3m_1m_2M}{c^5}\frac{1}{(a r_p)^{3/2}r_p}~.
\end{align}
The merger time for a highly eccentric BBH with initial periastron distance $r_{p0}$ and semi-major axis $a_0$ can be estimated by
\begin{align}
&t_{\rm gw}\simeq \int_{r_{p0}}^{a_0} \frac{da}{\vert\dot{a}\vert_{e=1}}\nonumber\\
&\simeq 1.20\times 10^7{\rm yr}\left(\frac{30^2\cdot 60M_\odot^3}{m_1m_2M}\right)\left(\frac{r_{p0}}{0.01{\rm au}}\right)^{7/2}\left(\frac{a_0}{\rm au}\right)^{1/2}.
\label{eq:t_merge}
\end{align}
Most of its time will be spent at its high initial eccentricity, when its periastron distance $r_p$ and maximum GW frequency $f_p$ are nearly invariant.

If the LIGO BH mergers arise from an initially highly-eccentric BBH population, with a given $a_0$ and $r_{p0}$, assuming that the population is in equilibrium, then it will have a large peak in GW frequency near the initial $f_{p0}=2 \sqrt{GM/(4\pi^2 r_{p0}^3})$. The indicated (density) distribution of systems near this frequency for the eccentric case should be much higher than that for the circular case:
\begin{align}
\frac{(dN/d\log_{10}f)_{\rm ecce}}{(dN/d\log_{10}f)_{\rm circ}} = \frac{\dot{f}_{\rm circ}}{\dot{f}_{p,\rm ecce}} &= \frac{96}{295}2^{9/2}\left(\frac{a_{0,\rm ecce}}{r_{p0}} \right)^{3/2}~.\nonumber\\ 
&\simeq 7400\epsilon_{0,\,-2}^{-3/2},
\label{eq:enhancement}
\end{align}
where $\epsilon_{0,\,-2}=(1-e_0)/10^{-2}=r_{p0}/a_{0,\rm ecce}/10^{-2}$, and where we have added the subscript ``ecce" to denote the initial properties of the highly eccentric binary. 

Note that the frequency distributions are different from the histograms of the number of systems binned in frequency. The ratio of the number of systems in the eccentric and circular cases near $f_{p0}$ can be estimated by the ratio of merger times (since most of merger time is spent near the initial orbital separation and frequency for both cases)
\begin{align}
    \left.\frac{N_{\rm ecce}}{N_{\rm circ}}\right\vert_{f_{p0}} &\sim \left.\frac{t_{\rm gw,ecce}}{t_{\rm gw,circ}}\right\vert_{f_{p0}}\simeq \frac{\int_{r_{p0}}^{a_0}da/\vert\dot{a}\vert_{e=1}}{\int_{0}^{r_{p0}}da/\vert\dot{a}\vert_{e=0}} \nonumber \\
    &= \frac{3}{425}2^{23/2}\left(\frac{a_{0,\rm ecce}}{r_{p0}} \right)^{1/2}=200\epsilon_{0,-2}^{-1/2}~.
\label{eq:number_ratio}
\end{align}

As an example, we take $m_1=m_2=30$\,M$_\odot$, $a_0=1\,$au, and $r_{p0}=0.008$, $0.015$, $0.030$, $0.068$\,au, respectively, and calculate $\dot{f}_p$ as a function of $f_p$. Assuming each system represents a population of migrating binaries and that each population makes up the entire observed LIGO rate, we obtain the equilibrium distribution as $dN/d\log_{10}f=\Gamma\ln(10) f_p/\dot{f}_p$. This rate is normalized such that the total rate of mergers is equal to the observed LIGO BBH merger rate of
$52.9_{-27.0}^{+55.6}\,{\rm Gpc}^{-3}{\rm yr}^{-1}$ \citep{2017PhRvL.118v1101A,2017ApJ...851L..35A,2017PhRvL.119n1101A,2018arXiv181112907T}. For illustration, we normalize to $50\,{\rm Gpc}^{-3}{\rm yr}^{-1}$ throughout this paper. Since the Milky Way-size galaxy number density is roughly $0.01\,{\rm Mpc}^{-3}$, we have $\Gamma\sim 5\times 10^{-6}\,{\rm yr}^{-1}$ per galaxy.

In Figure \ref{fig:dndlogf_singles_logscale2}, we show the frequency distribution (left) and the number histogram (right) for the circular case (blue) compared to the eccentric sample populations (orange to purple, respectively). The dots in the left panel denote the starting position of each population. The implied enhancement in the equilibrium density of systems in the eccentric case relative to the circular case is very large, in accord with equation (\ref{eq:enhancement}). Indeed, for $r_{p0}=0.008$\,au, the ratio of the eccentric population to the circular population is $>10^4$ at $\simeq0.7$\,mHz. However, the very highly-peaked enhancement for individual system starting parameters shown on the left panel becomes more modest when we compute the number per frequency bin, as shown on the right panel, because eccentric systems spend most of their time radiating at a small range of GW frequency.\footnote{Note that the height of the histogram depends on the bin spacing and the enhancement ratio in each bin is higher than estimated by Eq.~(\ref{eq:number_ratio}), because Eq.~(\ref{eq:number_ratio}) assumes most of the systems are near $f_{p0}$, but the time a system (circular or eccentric) spent within one bin is much shorter than its merger time.} In addition, as we show below, a more realistic eccentric BBH population is more broadly distributed in frequency by the realistic joint distribution of $(a_0,r_{p0})$ provided by any given formation scenario. As we show below, these factors reduce the magnitude of the eccentric-to-circular enhancement, but equations (\ref{eq:enhancement}) and (\ref{eq:number_ratio}) show that it is generic for any secularly-evolving eccentric merging BBH population that contributes at order unity to the observed LIGO rate.

\section{Results for Populations}
\label{sec:results}

In this section we give results for the equilibrium number of eccentric BBH systems in the Galaxy for several progenitor populations. 

\subsection{Simplest Population}
\label{subsec:simple}

To illustrate the scalings from Section \ref{sec:circ_ecce} we assume a generic eccentric BBH population motivated by dynamically-formed systems in dense stellar environments and few-body systems undergoing LK oscillations. We assume equal mass BBHs with $m_1=m_2=30\,M_\odot$, an initial semi-major axis distribution that is log-uniform between 1 and 1000$\,$au, and a thermal eccentricity distribution \citep{1919MNRAS..79..408J}, \ie, $e^2$ is uniform in [0,1]. Only systems with $t_{\rm gw}$ less than the Hubble time are included, since only these will contribute to the observed merger rate. The thermal eccentricity distribution is motivated by the properties of dynamically-formed binaries produced in dense stellar systems (e.g., \citealt{Samsing_DOrazio}), and by the fact that it produces a a uniform distribution of $r_{p0}$ in the $e\rightarrow 1$ limit, equivalent to a uniform distribution of the angular momentum squared $J^2$, which is a natural consequence of non-secular stochastic angular momentum kicks due to the tertiary in hierarchical triple systems \citep[\eg,][]{2012arXiv1211.4584K}.

\begin{figure}
\centering
\includegraphics[width=\linewidth]{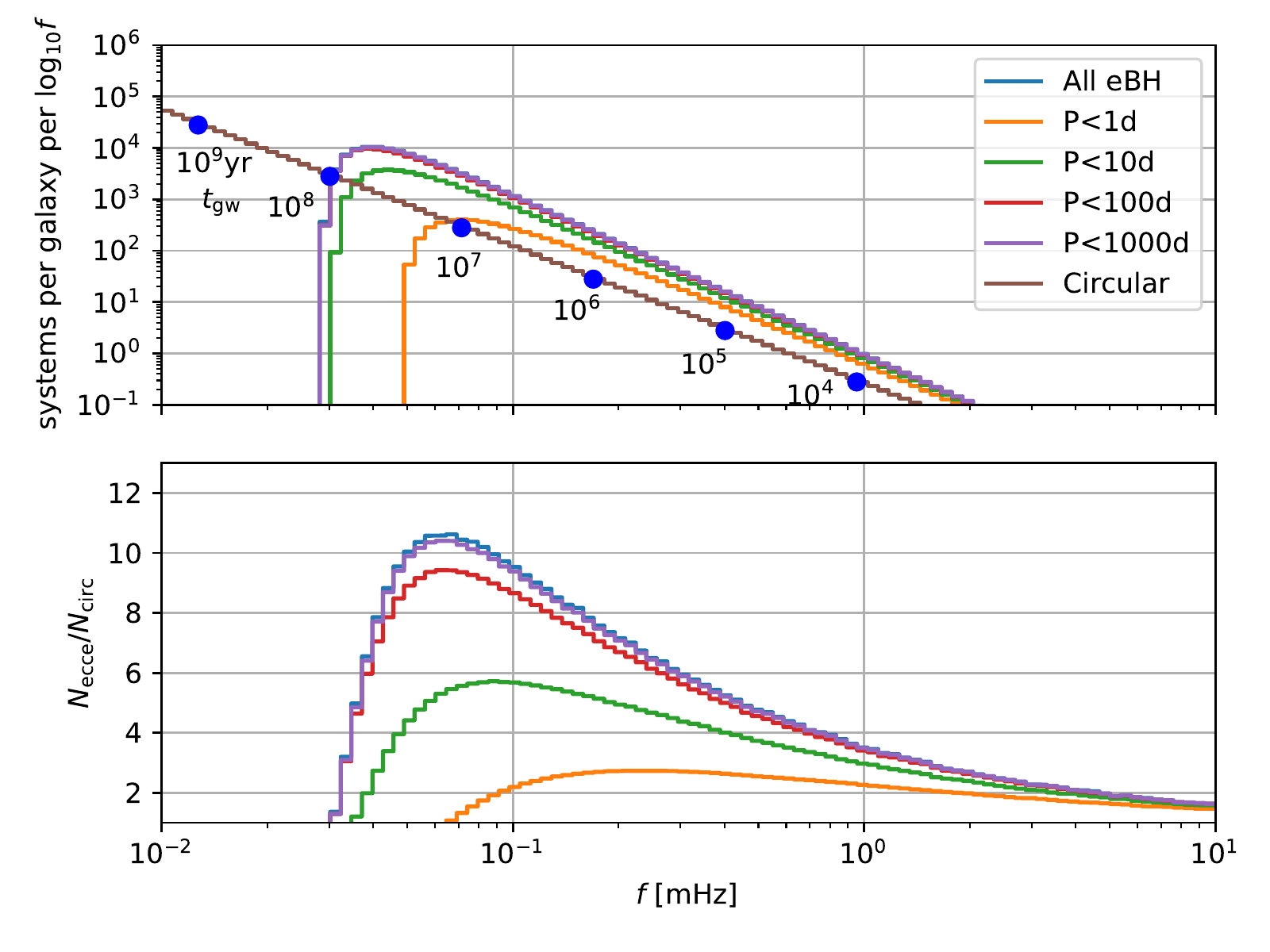}
\caption{\textit{Upper:} The peak frequency histograms of the simplest population of eccentric BBHs described in Section \ref{subsec:simple}, normalized to per $\log_{10}f$ bin. Solid curves show the distributions of all the eccentric BBHs (labelled ``All eBH") and those with orbital period $P<1$, $10$, $100$,and $1000$\,days. The distribution for circular BBHs is also shown. Both the eccentric and  circular channels are normalized to a merger rate of $\Gamma=5\times 10^{-6}$yr$^{-1}$ per Milky Way-size galaxy. Blue dots show the corresponding merger time $t_{\rm gw}$ of circular BBHs at each frequency. \textit{Lower:} Ratio of the number of eccentric systems to circular systems in each bin of frequency. An enhancement of $2-10$ is seen in frequency range $\simeq0.04-3$\,mHz.}
\label{fig:simplest}
\end{figure}

We exclude systems from our sample with 
\begin{equation}
    r_p<6.4\times 10^{-5}{\rm au}\left(\frac{m_1m_2a}{30^2M_\odot^2{\rm au}}\right)^{2/7}\left(\frac{M}{60M_\odot}\right)^{1/7}~,
    \label{highe_triple}
\end{equation}
in the $e\rightarrow 1$ limit, because the fractional change in the orbital energy per orbit is of order unity and the secular equations break down (i.e., $t_{\rm gw}\leq P$). Such systems emit GWs at a peak frequency of
\begin{equation}
    f_p>0.95\,{\rm Hz} \left(\frac{30^2M_\odot^2{\rm au}}{m_1m_2a}\right)^{3/7}\left(\frac{M}{60M_\odot}\right)^{2/7}.
\end{equation}
These extreme systems are not of relevance for the main comparison in the LISA band from $0.1-10$\,mHz, but may arise in nature  \citep{2017ApJ...836...39S, Samsing_DOrazio, 2018arXiv181111812K}.

We evolve any given binary system ``$i$'' in the sample using the secular equations and calculate $\dot{f}_p^{(i)}$ as a function of $f_p$. We assume each system represents an equilibrium population with the same initial conditions undergoing migration toward coalescence. Thus, at any peak frequency, the population $i$  has a distribution $dN^{(i)}/df_p\propto 1/\dot{f}_p^{(i)}$, and the total number of systems has a frequency distribution $dN/df=\sum_{i}dN^{(i)}/df_p=(\Gamma/N)\sum_i[1/\dot{f}_p^{(i)}]~$, where $N$ is the sample size, and the distribution is normalized to the LIGO rate of $50\,{\rm Gpc}^{-3}{\rm yr}^{-1}$. To obtain the number of systems $N_{12}$ in frequency bin $[f_1,f_2]$, we integrate the frequency distribution over the bin, $N_{12}=(\Gamma/N)\sum_it^{(i)}_{12}$, where $t^{(i)}_{12}$ is the time spent in the frequency bin for population $i$. 

Figure \ref{fig:simplest} shows the histograms of systems in the frequency range $[0.01,10]$\,mHz assuming that all the LIGO mergers come from either the eccentric channel or the circular channel. The upper panel shows the number of systems per logarithmic frequency bin, with the eccentric systems broken into sub-samples by orbital period. For the BBH population we consider here, when $a>11.5\,$au, the orbital period exceeds 5 years, which is roughly the operation timescale of LISA. Like single-transit planet detections in transit surveys (e.g., \citealt{Villanueva}), only a single pulse may be seen during the mission. However, systems with orbital periods of days or months will provide many repeated pulses during the entire mission (\S\ref{subsec:detection}).

The bottom panel of Figure \ref{fig:simplest} shows the ratio of the number of eccentric systems to the number of circular systems in bins of frequency. Note that this ratio does not depend on the overall normalization of the LIGO rate. We find $\simeq 6-10$ times more BBHs from the eccentric channel than that predicted by the circular channel at around $0.1$\,mHz, decreasing to $\simeq2$ times more at $\simeq3$\,mHz. In absolute numbers, we find that $\simeq45$, $90$, and $116$ eccentric BBHs in our Galaxy are currently emitting in the $0.1-1$\,mHz range with orbital periods $P\leq1$, $10$, and $100$\,days, respectively. These numbers are $\simeq2.5$, $5$, and $6.5$ times more than that predicted from the circular case in the same frequency band.

The distribution of BBHs with frequency, and thus the enhancement with respect to the circular channel, depend on the initial distribution of $(a_0,r_{p0})$. For more realistic estimates, in Sections  \ref{subsec:triple} and \ref{subsec:gc} we recompute the equilibrium distributions for eccentric BBHs arising from triple systems and dynamical interactions in globular clusters, respectively.

\subsection{Distributions from Triple Systems}
\label{subsec:triple}

Binaries in hierarchical triple systems are subject to gravitational perturbations from their tertiaries, and can be driven to high eccentricities due to the LK mechanism \citep{1962P&SS....9..719L,1962AJ.....67..591K}. In secular calculations where both the inner and outer orbits are averaged, the time over which the angular momentum of the inner binary is changed by order unity by the tertiary (the instantaneous LK timescale), in the $m_2\rightarrow 0$ limit, is given by \citep[\eg,][]{2014MNRAS.438..573B,2015MNRAS.452.3610A}
\begin{equation}
    t_{\rm LK}^{\rm (ins)}\sim\frac{8\sqrt{2}}{15\pi}\left(1+\frac{M}{m_3}\right)\frac{P_{\rm out}^2}{P}(1-e_{\rm out}^2)^{3/2}\sqrt{\frac{r_p}{a}}~,
\end{equation}
where $m_3$ is the tertiary mass, and $e_{\rm out}$ and $P_{\rm out}$ are the eccentricity and the outer orbital period, respectively.

The equilibrium  argument in Section \ref{sec:circ_ecce} relies on the assumption of dynamically isolated binaries whose frequencies evolve monotonically as a result of GW emission. This is only true for triple systems whose inner binaries are dynamically decoupled from the outer tertiary. A criterion for decoupling is that the inner binary is driven to sufficiently high eccentricity that $t_{\rm LK}^{\rm (ins)}$ becomes longer than $t_{\rm gw}$ or the General Relativistic (GR) precession timescale $t_{\rm prec}\sim 2c^2a^{3/2}r_p/[3(GM)^{3/2}]$.

In order to make a first estimate of the BBH population produced by triple systems, we run a secular calculation for triple systems with masses $m_1=m_2=m_3=30$\,M$_\odot$. The eccentricities of the inner and outer orbits, $e$ and $e_{\rm out}$, are both sampled from a thermal distribution. The semi-major axis of the inner orbit $a$ is sampled from a log-uniform distribution in $[10,1000]$\,au, and the semi-major axis ratio of outer to inner orbit, $a_{\rm out}/a$ is sampled from a log-uniform distribution in $[10,1000]$. We discard systems with $a_{\rm out}(1-e_{\rm out})<10a$ to make sure the validity of the secular calculation, and discard systems with $a_{\rm out}>10^5\,$au since they are too wide to make up an important fraction of triple systems. The orientations of both the inner and outer orbits are sampled randomly. We turn on the quadrupole-order term in the Newtonian 3-body perturbing Hamiltonian, which leads to the LK effect, the 1PN term (GR precession) and the 2.5PN GW dissipation terms for the inner orbit. We run $10^7$ systems for 10\,Gyrs, and find that $\sim 1.14\%$ systems experience a decrease in their semi-major axis of order unity. These systems dynamically decouple from the tertiary and will merge within a relatively short time. We take the last eccentricity maximum of each such system and its semi-major axis $a$, and use them to set the initial conditions $(a_0,r_{p0})$ of ``isolated binaries'' in our calculation of the equilibrium distribution of the population.  

\begin{figure}
\centering
\includegraphics[width=\linewidth]{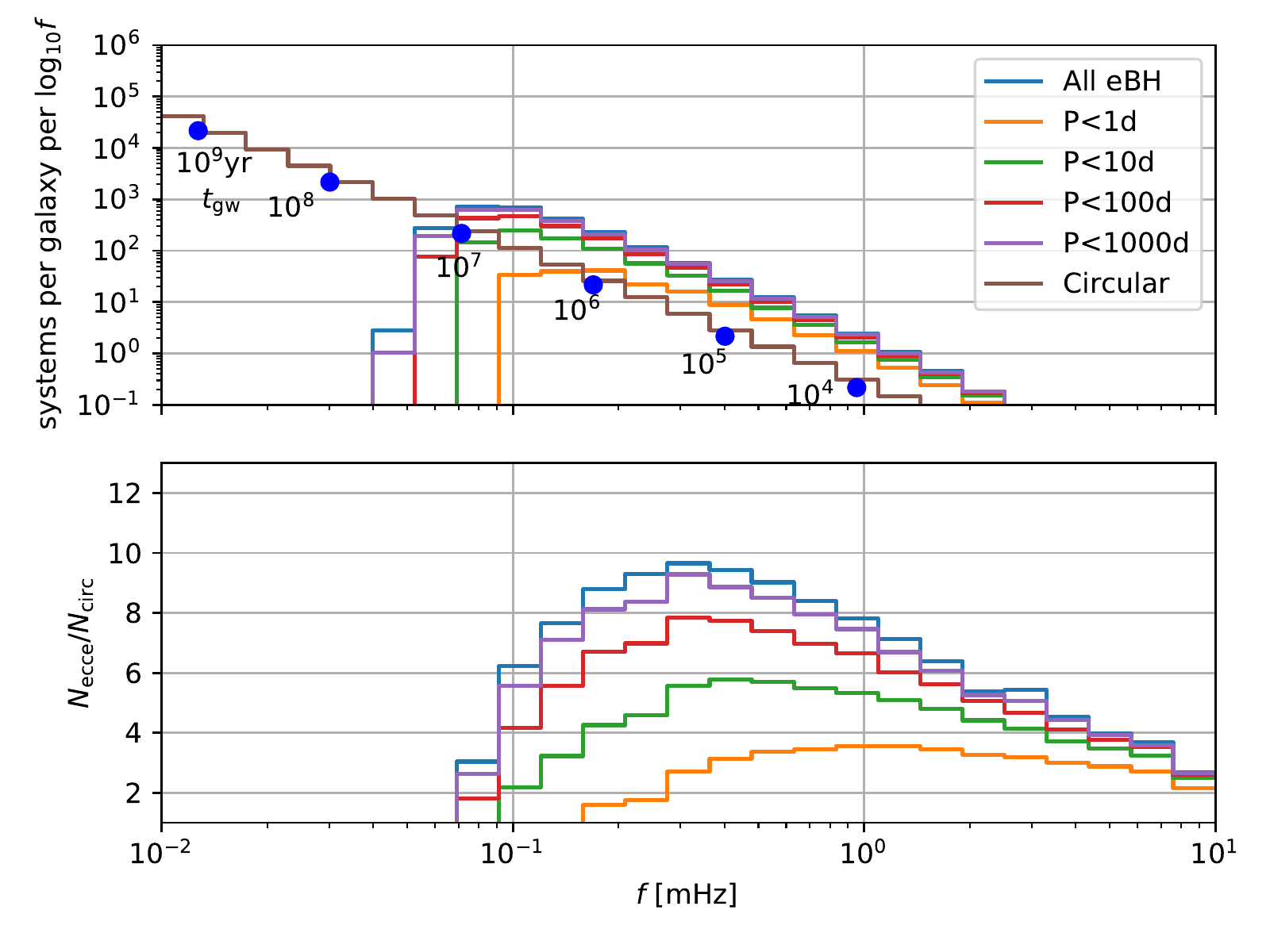}
\caption{Same as Figure \ref{fig:simplest}, but for the population of eccentric BBHs from secular dynamics of hierarchical triple systems described in Section \ref{subsec:triple}. \textit{Upper:} Peak frequency histograms for eccentric an circular BBHs. \textit{Lower:} Ratio of eccentric  to circular systems. A significant enhancement is seen in frequency range $0.1-10$\,mHz. The total number of eccentric systems with orbital periods of days to months is expected to be of order $10-100$.}
\label{fig:triple1e7}
\end{figure}

Following the same procedure as in Section \ref{subsec:simple}, we obtain the peak frequency histograms as shown in Figure \ref{fig:triple1e7}. We again find a significant enhancement of the eccentric BBH population in frequency range $0.1- 1$\,mHz, in which the absolute number of systems with orbital periods within $P\leq1$, $10$, and $100$\,days is $\simeq20$, $80$, and $130$. Comparing to Figure \ref{fig:simplest}, the systems at $f_p\lesssim0.05$\,mHz disappear. This comes from the fact that such systems still have perturbations from their tertiaries and are thus not dynamically isolated. However, the enhancement between 0.1 and 1 mHz persists, and is a factor of $\simeq10$ at $\sim0.3$\,mHz.

Note that many triple systems undergoing LK oscillations may emit GWs in the $\simeq0.1-10$\,mHz band, but may not be dynamically decoupled by our criterion. Such systems are not included here because they do not obey the equilibrium assumptions set out in Section \ref{sec:circ_ecce}. The total population of GW emitters in the $0.1-10$\,mHz band (whether dynamically decoupled or not) has yet to be computed for a realistic and evolving distribution of massive triple systems as the Galaxy forms over cosmic time.

Figure \ref{fig:triple1e7} gives just one minimal estimate for the distribution from triple systems. Different component binary masses, which lead to octupole-order terms in the 3-body Hamiltonian, tertiary masses, initial orbital parameter distributions, and cuts on the resulting population can quantitatively affect the results. Several variations are presented in Appendix \ref{app:tests}, with maximum and minimum enhancements relative to the circular case of $\simeq2-15$.

\subsection{Distributions from Globular Clusters}
\label{subsec:gc}

BBHs arising in globular clusters (GCs) and other dense stellar environments provide another important channel for eccentric BH migration. Recent numerical studies show that three populations of BBHs are produced during few-body scattering in GCs. One is produced by chaotic 3-body motion, leading to BBH mergers in the cluster at very high eccentricity such that $f_p\gtrsim 1$\,Hz and $t_{\rm gw}$ becomes less than the orbital period $P$ (as in eq.\ \ref{highe_triple}). These systems evolve dynamically and never enter the $0.1-1$\,mHz band. A second physical class of mergers are those from BBHs excited to high enough eccentricity within the cluster that $t_{\rm gw}$ becomes shorter than the time between two interactions. The third class is those BBHs ejected from the cluster. Adopting the nomenclature of \citep{Samsing_DOrazio} we refer to these three classes as ``3-body mergers," ``2-body mergers," and ``ejected mergers," respectively. The latter two classes evolve secularly through the $0.1-1$\,mHz band and are dynamically isolated \citep{Samsing_DOrazio,2018arXiv181111812K}.

We set aside the dynamically merging 3-body mergers and consider only 2-body mergers and ejected mergers. As an illustration, for these two physical categories, we adopt the distribution of BBH parameters resulting from the semi-analytic model described in \cite{Samsing_DOrazio}. The binary component masses are assumed equal with $m=30M_\odot$. The semi-major axis and eccentricity distributions for the BBHs when they are dynamically isolated are given by \cite{Samsing_DOrazio}, and are used to set the initial conditions of ``isolated binaries'' in our calculation of the equilibrium BBH distribution, just as in Section \ref{subsec:simple}.

\begin{figure}
    \centering
    \includegraphics[width=\linewidth]{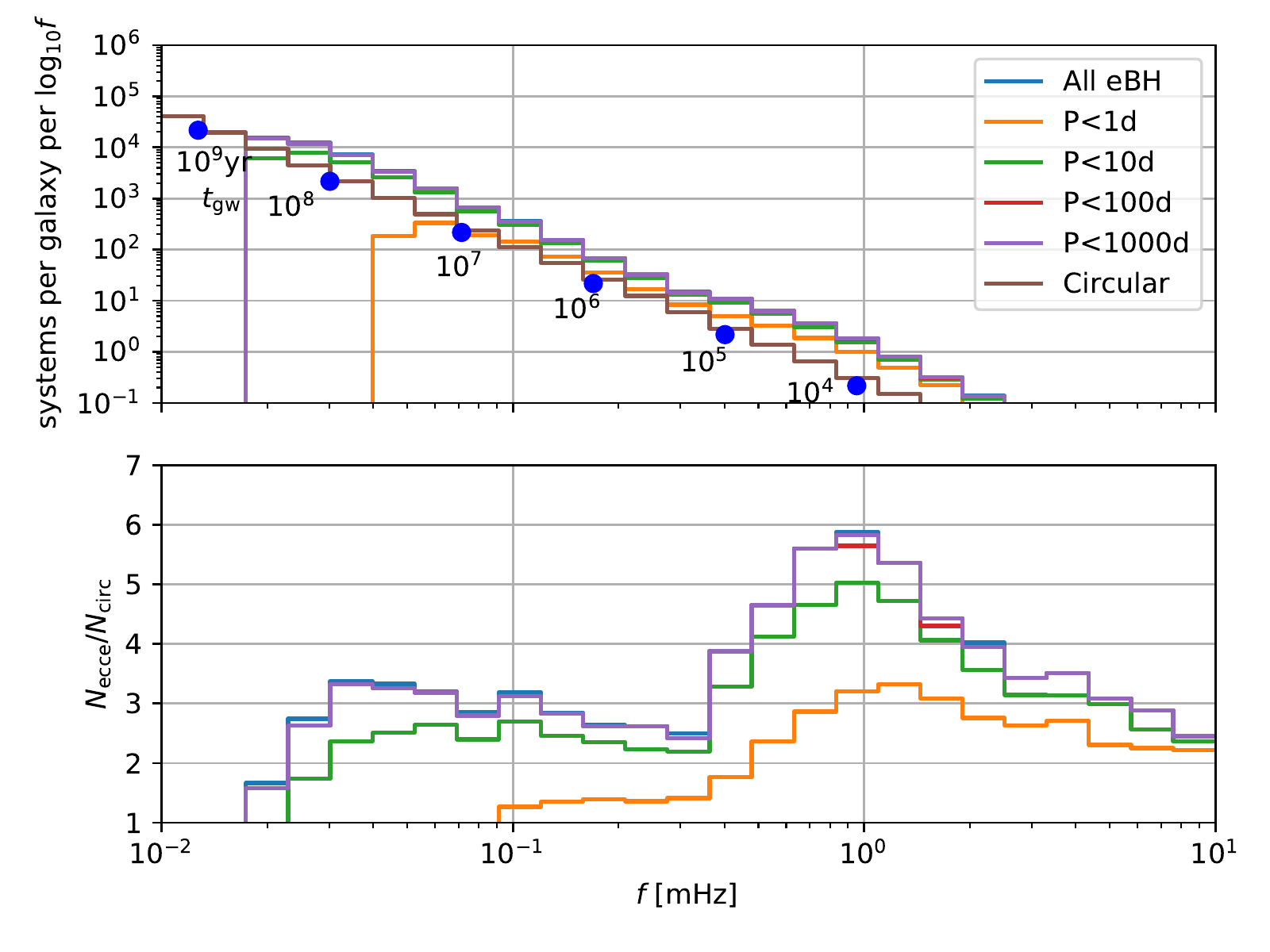}
    \caption{Same as Figures \ref{fig:simplest} and \ref{fig:triple1e7}, but for the population of eccentric BBHs from GCs described in Section \ref{subsec:gc}. \textit{Upper:} Peak frequency histograms for eccentric and circular BBHs. \textit{Lower:} Ratio of eccentric to circular systems. The ``2-body" BBH mergers inside GCs dominate the peak in the distribution at $\simeq 0.4-10$\,mHz, while the ``ejected" BBH mergers contribute over a broader frequency range \citep{2018arXiv181111812K}.}
    \label{fig:dndlgf_gc1}
\end{figure}

Figure \ref{fig:dndlgf_gc1} shows the peak frequency histogram (top) and the ratio of eccentric to circular systems (bottom) for dynamically-formed eccentric BBHs, normalized to the LIGO rate, as in previous figures. The shape of the number of systems per bin encodes the formation channel. As more clearly shown in the ratio plot (bottom), the 2-body BBH mergers inside GCs dominate the distribution from $0.4-10$\,mHz, producing a peak relative to the circular case at $f_p\gtrsim 1\,$mHz. The ejected BBH mergers resulting from binaries kicked out of GCs contribute to a much wider range of frequencies, with a low-frequency cut-off at $\simeq0.02$\,mHz for the $P\gtrsim10$\,day binaries.  

\section{Discussion and Conclusion}
\label{sec:discussion}
Assuming the distribution of BBHs is in equilibrium, producing a steady merger rate as seen by LIGO, eccentric BBH formation channels predict a significantly different population distribution in GW frequency than for circular BBH formation channels. Because eccentric BBHs spend more time radiating in the $0.1-1$\,mHz GW band they should generically outnumber circular systems at the same frequency. We estimate that there are $\sim 10-100$ systems with GW peak frequencies of $0.1-1$\,mHz in our Galaxy, which is $\simeq2-10$ times higher than that predicted for circular BBHs. Dozens of the eccentric systems will have orbital periods of order days to months, implying they may be detectable.

\subsection{Detectability}
\label{subsec:detection}

To estimate the two-detector sky-averaged signal-to-noise ratio (SNR) for eccentric BBHs, we start from Eq.~(45) in \cite{2004PhRvD..70l2002B}, \ie, summing over contributions from all harmonics of the orbital frequency $f_{\rm orb}$,  
\begin{equation}
    {\rm SNR}^2 = 2 \sum_{n=1}^{n_{\rm max}}\int \frac{h_{c,n}^2}{f_n S_h(f_n)}d\ln f_n~,
\label{eq:snr}
\end{equation}
where $n_{\rm max}$ is the maximum harmonic used in fitting, $f_n\equiv nf_{\rm orb}$, $S_h$ is the full strain spectral sensitivity density including the LISA instrumental noise and the confusion noise from the unresolved galactic binaries \citep[\eg][]{2018arXiv180301944R}\footnote{Note that the pre-factor ``2'', due to the fact that LISA has two channels, has been absorbed into $S_h$ in \cite{2018arXiv180301944R}.}. The characteristic amplitude $h_{c,n}$ is given by $h_{c,n} = (\pi D)^{-1}\sqrt{2\dot{E}_n/\dot{f}_n}$, where the unit $G=c=1$ has been applied. $D$ is the distance of the source. $\dot{E}_n$ is the GW energy emission rate in the $n$-th harmonic and is given by
\begin{equation}
    \dot{E}_n = \frac{32}{5}M_c^{10/3}(2\pi f_{\rm orb})^{10/3}g_n(e)~,
\label{eq:dotEn}
\end{equation}
where $M_c\equiv (m_1m_2)^{3/5}M^{-1/5}$ is the chirp mass (for $m_1=m_2=30M_\odot$, $M_c=26M_\odot$), $g_n(e)$ is given by Eq.~(20) in \cite{1963PhRv..131..435P}. Since the eccentric migration timescale ($\sim t_{\rm gw}$) is much longer than the mission lifetime $\tau$, the orbital frequency does not change much during the mission, the integral in the SNR becomes
\begin{equation}
    \int \frac{h_{c,n}^2}{f_n S_h(f_n)}d\ln f_n \simeq \frac{h_{c,n}^2}{f_n S_h(f_n)}\frac{\dot{f}_n \tau}{f_n}=\frac{2\dot{E}_n\tau}{\pi^2 D^2 f_n^2 S_h(f_n)}.
\end{equation}
Combining it with Eq.~(\ref{eq:dotEn}) we obtain 
\begin{equation}
    {\rm SNR}^2 = \frac{512\tau}{5D^2}\frac{(GM_c)^{10/3}}{c^8}(2\pi f_{\rm orb})^{4/3}\sum_{n=1}^{n_{\rm max}}\frac{g_n(e)}{n^2S_h(f_n)}~.
\end{equation}
As discussed by \cite{2011ApJ...729L..23G} in the context of eccentric binary white dwarfs, the GW emission is dominated by pulses at periastron. 

\begin{figure}
    \centering
    \includegraphics[width=\linewidth]{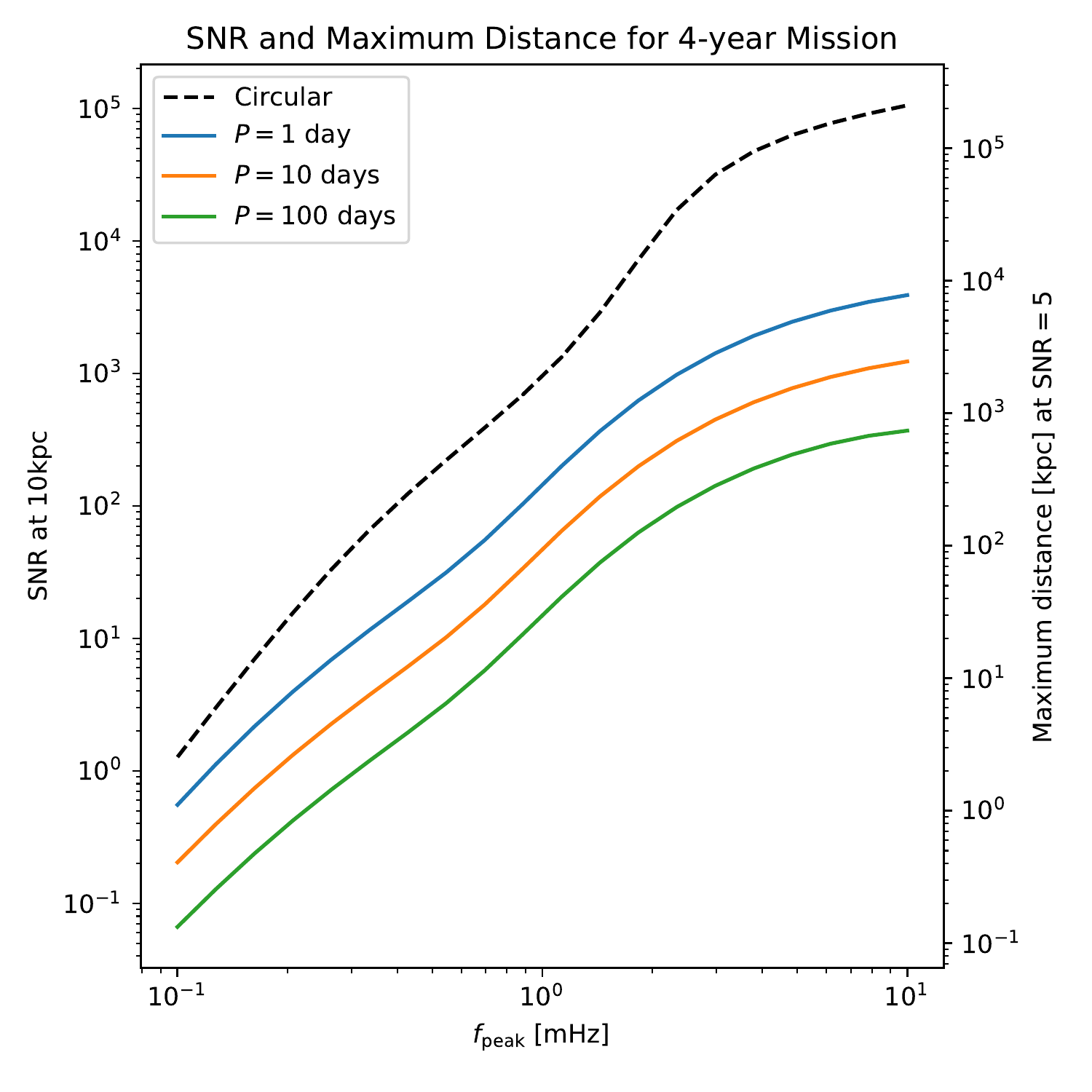}
    \caption{The left $y$-axis reads the estimated SNRs for detecting eccentric BBH systems with $M_c=26M_\odot$, $P=1,10,100\,$days, locating at 10kpc, in a 4-year mission. The SNR scales proportionally with $\sqrt{\tau}/D$. The right $y$-axis reads the maximum distance for detecting these systems requiring SNR$=5$. Note that the cosmological redshift is neglected, which only lowers the frequencies by 0.23\% at 10$\,$Mpc. For comparison, we also show the detectability of circular BBHs in the dashed line.}
    \label{fig:snr}
\end{figure}

Taking $\tau=4\,$years, $D=10\,$kpc, $n_{\rm max}=10^5$, $f_p=0.5\,$mHz, $M_c=26M_\odot$, for an eccentric BBH with $P=1$, $10$, and $100$\,days, we obtain ${\rm SNR}=26$, $8.5$, and $2.7$, respectively, implying that a number of eccentric BBHs will be detectable by LISA. In Figure \ref{fig:snr}, we show the SNRs for systems at 10\,kpc with different peak frequencies and $P=1$, $10$, and $100$\,days. Requiring ${\rm SNR}=5(2)$ for detection and assuming a distance of 10\,kpc (which gives a conservative estimate of the number of observable systems), we estimate that $\sim 7$, $7$, and $8$ (15, 17, and 17) eccentric BBHs with $P$ less than 1, 10, and 100\,days will be detected in our Galaxy in $0.1-1\,$mHz in the case of the simplest distribution, while $\sim 5$, $6$, $7$ (11, 14, 15) systems in the triple case, and $\sim 4$, $4$, $5$ (8, 9, 10) systems in the GC case. While the number of detectable eccentric systems is largely limited by the SNR at $f_p\lesssim 0.2\,$mHz, where many more systems exist, the detection of the few systems at higher frequencies immediately implies the existence of the eccentric BBH population. Also note that although less systems are present at higher frequencies in our Galaxy, a larger volume is accessible due to the larger SNRs, hence more extragalactic sources may be detectable \citep[\eg][]{Samsing_DOrazio,2018arXiv181104926R,2018arXiv181111812K}. These sources could also be interesting to experiments sensivite to slightly higher frequency bands, such as DECIGO, Taiji and TianQin. The right axis of Figure \ref{fig:snr} shows that eccentric systems with $P<1$\,day can be discovered by LISA out to $\sim8$\,Mpc distances.

In Table \ref{Tab:numbers}, we show the estimated numbers of LISA-detectable galactic eccentric and circular BBHs assuming different LIGO rates due to the uncertainty of the measured LIGO merger rate. Although the actual number of circular systems in each frequency bin is less than that of eccentric systems, the detectable numbers may be similar, due to better SNRs of detecting circular binaries. Note that the numbers are obtained by assuming all the systems are at distance 10$\,$kpc, a typical value of their average distance. While closer systems have a higher SNR, much less systems exist at closer distances if a uniform spatial distribution of BBHs in the stellar halo is assumed, hence resulting in little changes in the estimated numbers of observable systems.

\begin{table}
\centering
\begin{tabular}{| c | c | c | c |}
 \hline
 $\Gamma$ (Gpc$^{-3}$yr$^{-1}$) & 25 & 50 & 100 \\
 \hline\hline
 & \multicolumn{3}{c|}{Circular}\\
 \hline
 SNR$\ge 2$ & 8 & 16 & 33\\
 \hline
 SNR$\ge 5$ & 6 & 11 & 23 \\
 \hline\hline
 & \multicolumn{3}{c|}{Simplest ($P<1,10,100$ days)}\\
 \hline
 SNR$\ge 2$ & 8,9,9 & 15,17,17 &30,34,35 \\
 \hline
 SNR$\ge 5$ &3,4,4  & 7,7,8 &14,15,15 \\
 \hline\hline
 & \multicolumn{3}{c|}{Triple ($P<1,10,100$ days)}\\
 \hline
 SNR$\ge 2$ &5,7,7  &11,14,15 &22,29,30 \\
 \hline
 SNR$\ge 5$ &3,3,3  &5,6,7 &11,13,13 \\
 \hline\hline
 & \multicolumn{3}{c|}{GC ($P<1,10,100$ days)}\\
 \hline
 SNR$\ge 2$ &4,5,5 &8,9,10 &16,19,19 \\
 \hline
 SNR$\ge 5$ &2,2,2  &4,4,5 &8,9,9 \\
 \hline
\end{tabular}
\vspace*{0.3cm}
\caption{The estimated numbers of LISA-detectable Galactic BBHs for a 4-year mission and for SNR$\ge 2$ and 5, assuming all the LIGO BH mergers originate from the ``circular'' channel, the simplest eccentric BBH population, the triple scenario, or the GCs. We scale the numbers for different LIGO BH merger rates $\Gamma$ due to the uncertainties of the measured LIGO merger rate.}
\label{Tab:numbers}
\end{table}

\subsection{Complexities}
\label{subsec:complex}

A primary assumption in the results presented here is that the numbers of BBH systems is based on the equilibrium assumption, which will break down on $10$\,Gyr timescales. However, most of the eccentric migrating systems in triples and GCs with $f_p>0.1\,$mHz have merger times less than 1\,Gyr, during which the variations expected as a result of the time history of star formation in the Galaxy may not be significant. 

Additional uncertainties lie in the merger rate. There is a factor of a few uncertainty in the LIGO BBH merger rate, which is a function of BBH mass, and we have also assumed that the LIGO BBH merger rate applies to our Galaxy. While these uncertainties will affect the absolute numbers of systems expected, the ratios between the eccentric and circular cases presented are robust. In reality, the merger rate may be a mixture of all the possible channels. For the GC case, most of cluster simulations predict merger rates less than $50\,$Gpc$^{-3}$yr$^{-1}$ \citep[\eg,][]{2018ApJ...866L...5R}. Thus, the combined number of eccentric BBHs and their frequency distribution may depend on the fraction of each channel, which might in turn be used to probe the relative importance of various channels by LISA. 

For the case of eccentric BBHs produced in triple systems, there are a number of uncertainties and complexities. These include (1) octupole-order perturbations for a realistic population, (2) evection, and (3) the astrophysics of realistic triple system masses and their evolution. The octupole-order perturbation (1) arises when the inner binary has unequal masses \citep[\eg][]{2013MNRAS.431.2155N} and may change the frequency distributions as indicated in the tests in Appendix \ref{app:tests}. A more comprehensive analysis with a realistic BH mass distribution is needed to fully explore its effect. Evection (2) induces eccentricity oscillations of the inner orbit on timescale of $P_{\rm out}$, which may be important at high eccentricities where the inner orbit has small angular momentum, and is thus prone to torque from the tertiary \citep[\eg,][]{2005MNRAS.358.1361I,2012arXiv1211.4584K,2014MNRAS.439.1079A,2018MNRAS.476.4234F}. However, we neglect it here because evection may be considered as random kicks to the inner orbit and will produce a thermal distribution of $e$, which is already assumed in our initial distribution. Thus, while evection may change the absolute number of triple systems that produce merging and dynamically isolated BBHs, and while individual systems may experience non-secular changes of $r_p$, the overall distribution should not be modified by evection. The inclusion of evection may also introduce non-negligible secular effects when the triple systems are moderately hierarchical \citep[\eg,][]{2016MNRAS.458.3060L,2018MNRAS.481.4602L,2018MNRAS.481.4907G}, which could suppress the octupole-order oscillations, and thereby affect the resulting distribution of systems as a function of GW frequency. Finally, (3) we neglect stellar evolution of the massive star progenitors, including possible mass transfer, adiabatic and dynamical mass-loss, and natal kicks due to the recoil during asymmetric supernova explosions. These effects may change the orbital parameter distributions or unbind the systems, which may suppress this formation channel \citep[\eg,][]{2017ApJ...836...39S}. Additionally, as Appendix A shows (see Run 3), a realistic distribution of tertiary masses can significantly change the relative enhancement of eccentric to circular systems.

\subsection{Summary}
Our major findings in this paper are as follows.
\begin{enumerate}
    \item Assuming the distribution of binary black holes (BBHs) is in equilibrium, producing a steady merger rate as seen by LIGO, we show that eccentric BBH formation channels predict a larger number of systems relative to circular BBH formation channels throughout the $0.1-1$\,mHz GW frequency band. Equations (\ref{eq:enhancement}) and  (\ref{eq:number_ratio}) and Figure \ref{fig:dndlogf_singles_logscale2} show that this predicted enhancement is generic, and follows from the fact that eccentric systems spend more time radiating in the low-frequency GW band than circular systems.
    
    \item We estimate the absolute number of radiating systems in the circular and eccentric cases in the Galaxy. Figures \ref{fig:simplest}, \ref{fig:triple1e7}, and \ref{fig:dndlgf_gc1} show the eccentric and circular cases for a generic eccentric population, for eccentric BBHs produced by triple systems undergoing Lidov-Kozai oscillations, and BBHs formed dynamically in globular clusters, respectively. Assuming that both eccentric and circular channels produce the observed LIGO rate, we find that eccentric systems outnumber circular systems by a factor of $2-10$ throughout the $0.1-10$\,mHz GW band. Under these assumptions, there are $\sim 10-100$ eccentric BBHs with GW peak frequencies of $0.1-1$\,mHz in the Galaxy. Dozens of these systems have orbital periods of order days to months.
    
    \item Eccentric BBH systems emit GW pulses at periastron. We calculate the signal-to-noise ratios (SNR) for detecting eccentric binaries with a LISA-like sensitivity curve, and estimate that $\simeq 7$ (15) eccentric systems should be seen in the Galaxy with SNR\,$\ge5$(2), slightly less than the number of observable circular systems (11 and 16 for SNR$\,\ge\,$5 and 2) in the range of 0.1-1$\,$mHz. See Figure \ref{fig:snr} and Table \ref{Tab:numbers}. For the rarer eccentric systems with higher peak GW frequency of $1-10$\,mHz, the detection volume increases to $0.8-8$\,Mpc for systems with orbital periods less than $\simeq1$\,day. 
    
\end{enumerate}

\acknowledgments
We thank Johan Samsing, Carl Rodriguez, Pierre Christian, Paulo Montero for useful discussions and Kyle Kremer for very detailed and helpful comments on the manuscript. We are also grateful for suggestions from an anonymous referee which improved the paper. XF is supported by the Simons Foundation and NSF 1313252. TAT is supported in part by NSF 1313252, an IBM Einstein Fellowship from the Institute for Advanced Study, Princeton, and a Simons Foundation Fellowship. CMH is supported by the Simons Foundation, the US Department of Energy, the Packard Foundation, NASA, and the NSF. Many computations in this paper were run on the CCAPP condo of the Ruby Cluster at the Ohio Supercomputer Center \citep{OhioSupercomputerCenter1987}.

\appendix
\section{Tests of Different System Configurations}
\label{app:tests}
In this appendix, we carry out tests for triple systems with different masses, initial orbital parameter distributions and cuts. We summarize the runs (original and 5 tests) in Table \ref{Tab:tests}.

\begin{figure*}
\centering
        \subfigure[Run 0]{
                \centering \includegraphics[width=.5\linewidth]{triple1e7_4_average_dNdlgf_binned.pdf}
        }%
        \subfigure[Run 1]{
                \centering \includegraphics[width=.5\linewidth]{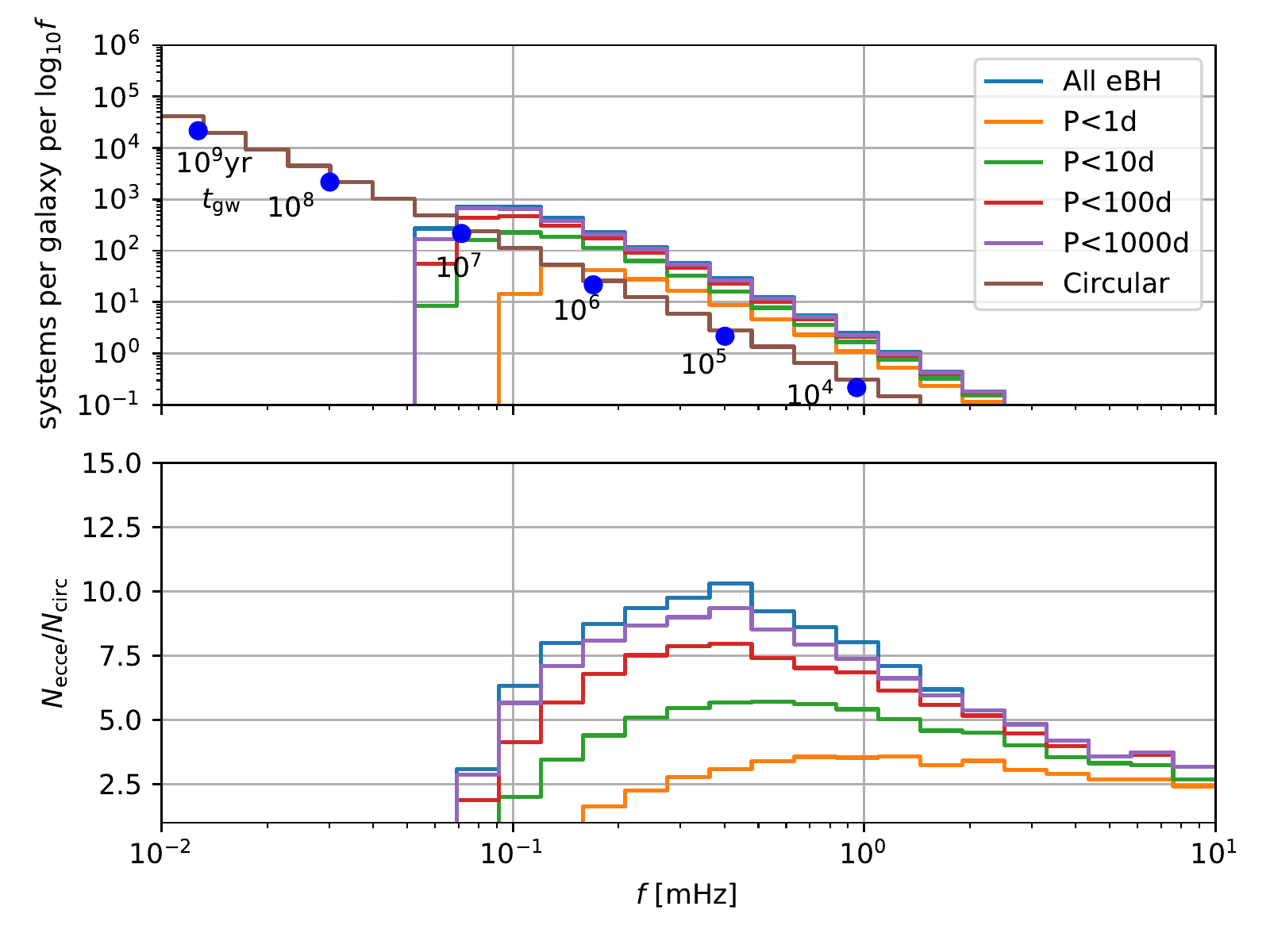}
        }%
        \\
        \subfigure[Run 2]{
                \centering \includegraphics[width=.5\linewidth]{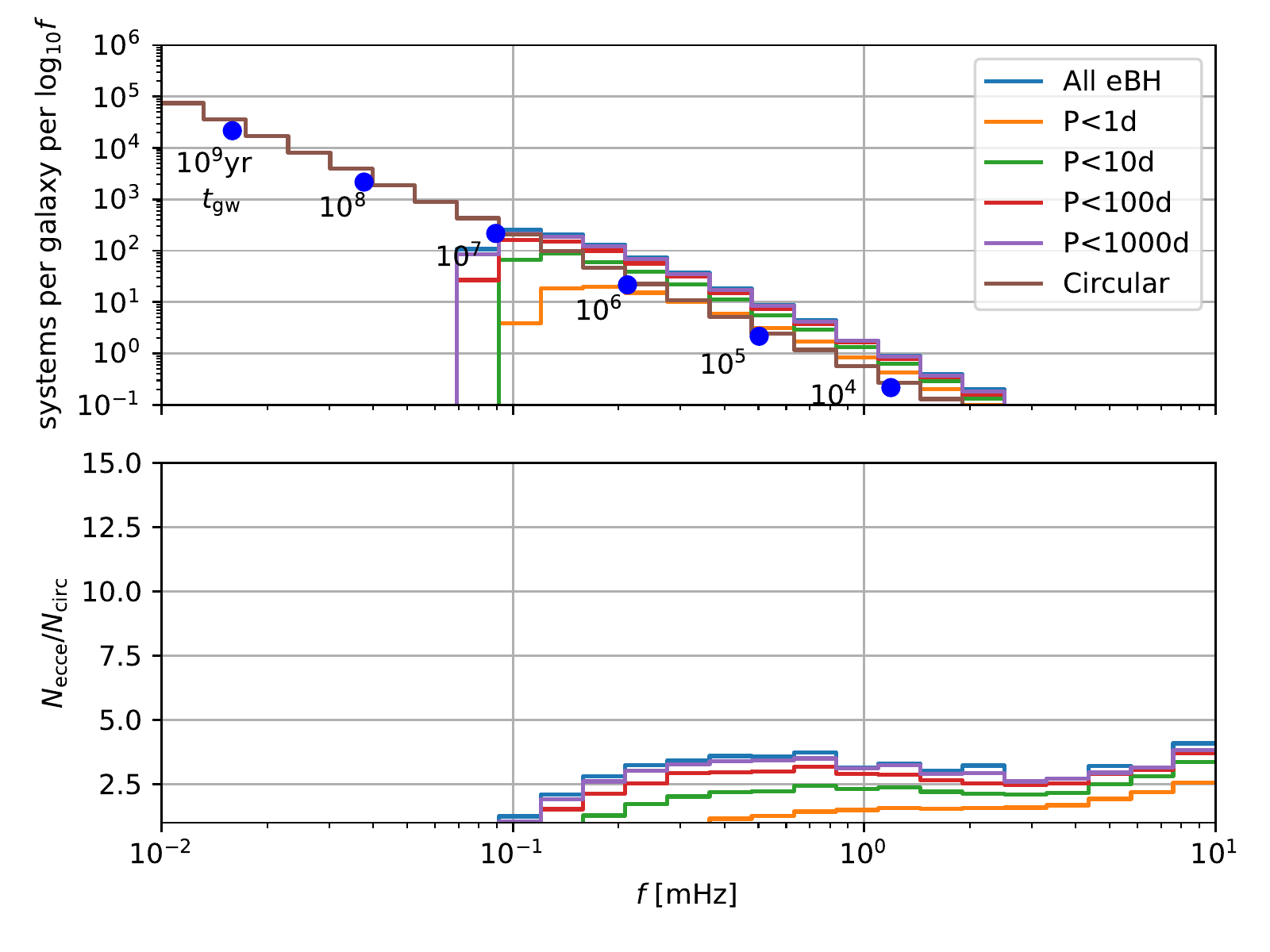}
        }%
        \subfigure[Run 3]{
                \centering \includegraphics[width=.5\linewidth]{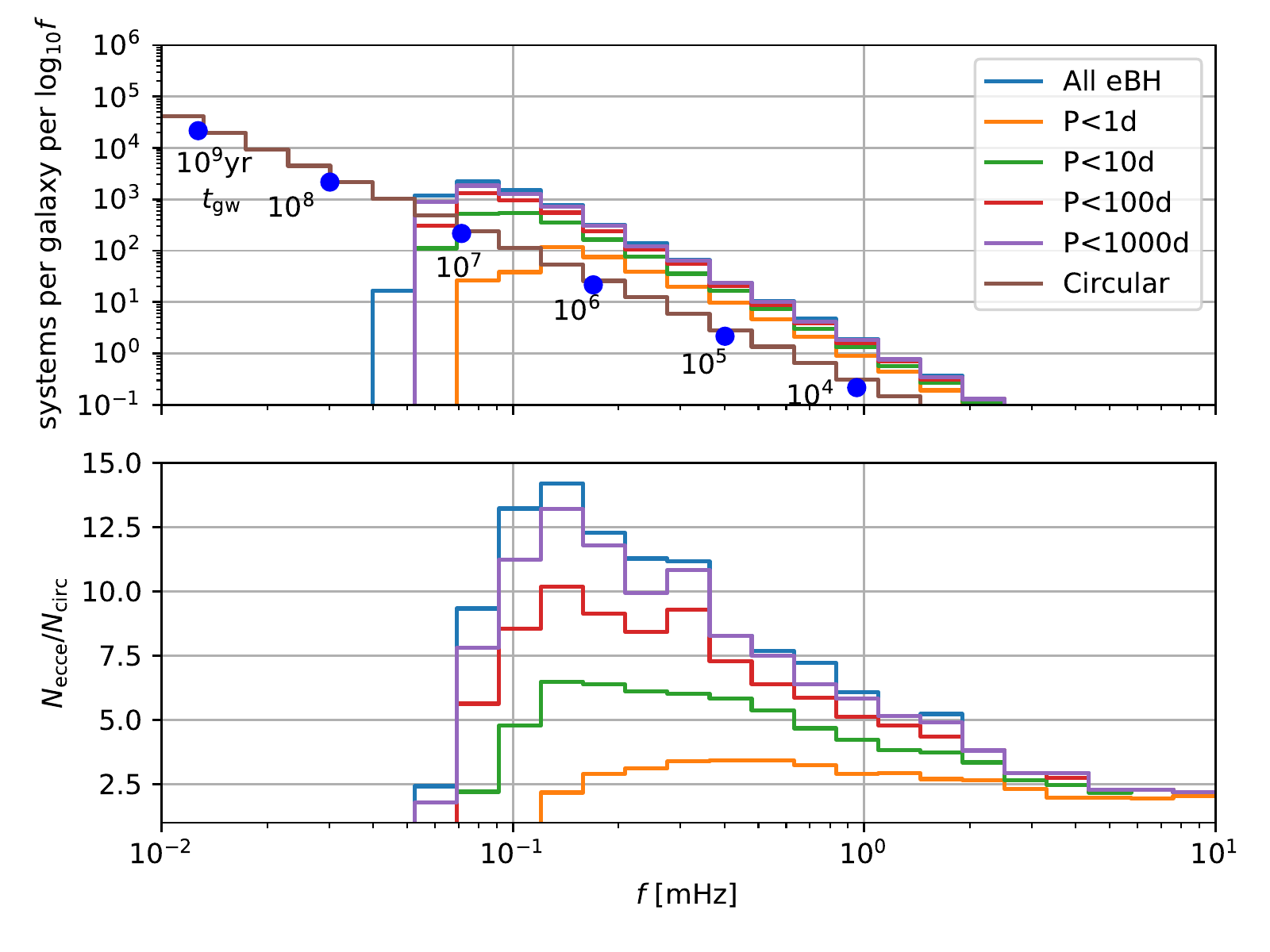}
        }%
        \\
        \subfigure[Run 4]{
                \centering \includegraphics[width=.5\linewidth]{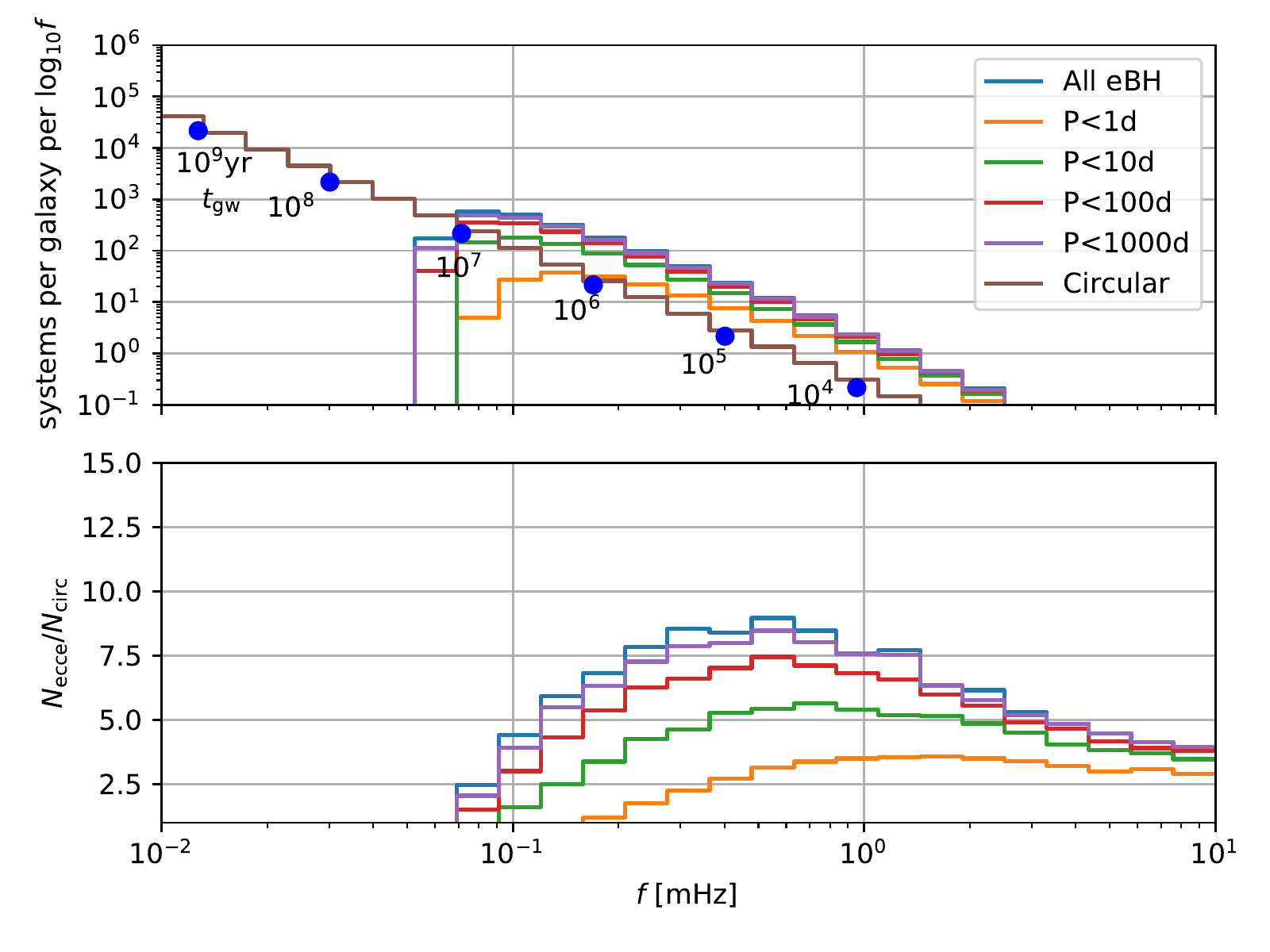}
        }%
        \subfigure[Run 5]{
                \centering \includegraphics[width=.5\linewidth]{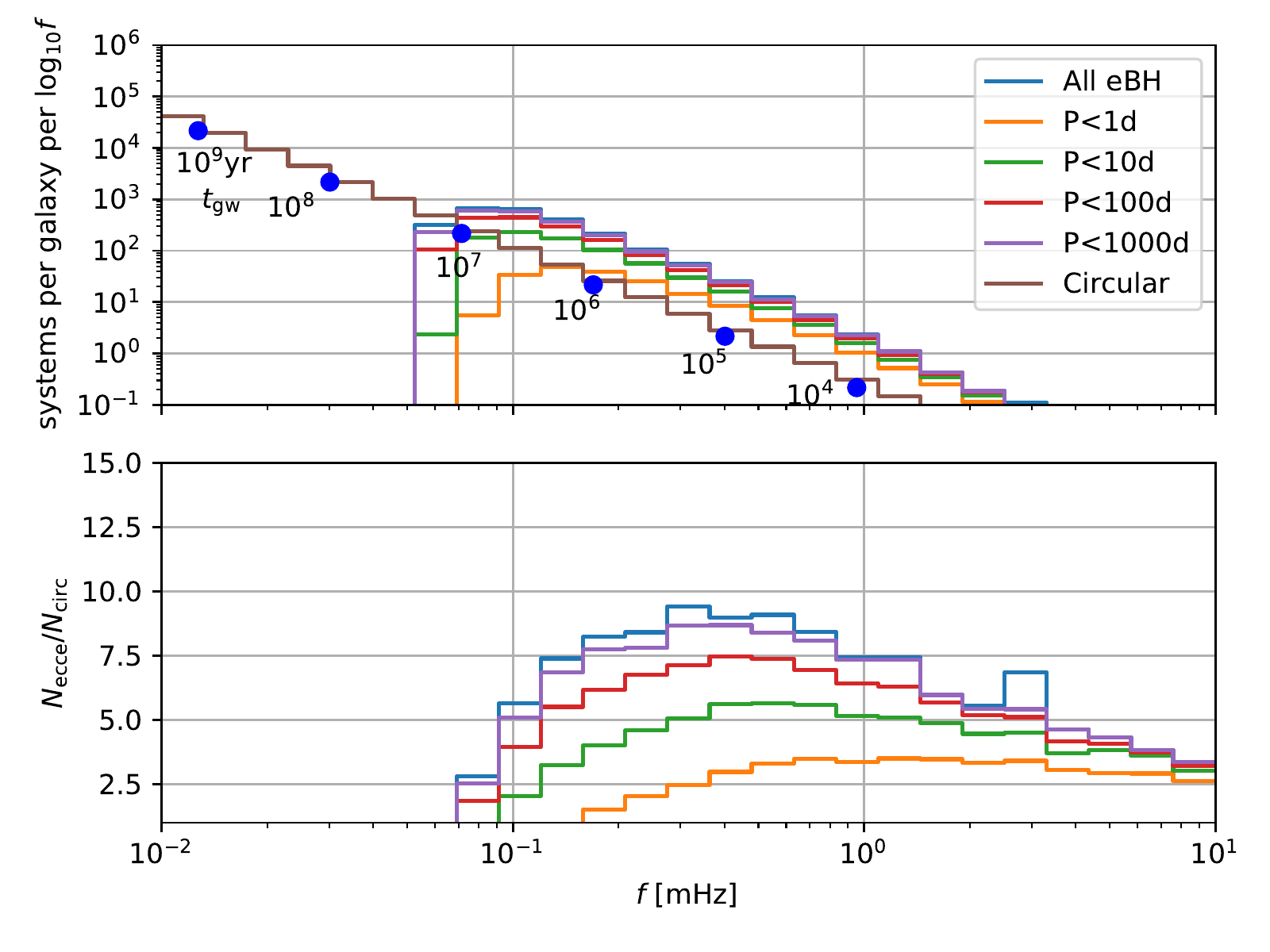}
        }%
        \caption{Peak frequency histograms from different runs listed in Table \ref{Tab:tests}, normalized to per $\log_{10}f$ bin.}\label{fig:tests}
\end{figure*}

For each run, we use the orbital parameters of the mergers to produce the peak frequency histograms. The results are shown in Figure \ref{fig:tests}. All the runs except Run 2 have shown consistent enhancements in the 0.1-1 mHz frequency range. In Run 2, the curves move towards higher frequencies, because the inner binary mass is smaller (30+15 $M_\odot$), leading to a longer merger time, hence requiring a smaller initial periastron (larger initial peak frequency) to merge within 10 Gyr. Meanwhile, the octupole-order perturbation in the triples drives many systems to very high eccentricities (high $f_p$), as seen in Figure \ref{fig:tests}(c). In addition, the $\dot{f}$ for the corresponding circular case is smaller, resulting in a larger $dN/d\log_{10}f$ value. All the factors reduce the enhancement in the 0.1-1 mHz range.

\begin{table}
\centering
\begin{tabular}{| c | c | c | c | c | c | c |}
 \hline
 \multirow{2}{*}{Runs}& \multicolumn{3}{ c| }{masses ($M_\odot$)} & \multirow{2}{*}{$a$} & \multirow{2}{*}{cuts} & \multirow{2}{*}{merger fractions in 10 Gyr} \\ \cline{2-4} & $m_1$ & $m_2$ & $m_3$ & & & \\
 \hline
 0 & 30 & 30 & 30 & log-uniform & $a_{\rm out}(1-e_{\rm out})\geq 10a$ & $1.14\%$\\
 1 & 30 & 30 & 30 & log-normal & $a_{\rm out}(1-e_{\rm out})\geq 10a$ & $1.14\%$\\
 2 & 30 & 15 & 30 & log-uniform & $a_{\rm out}(1-e_{\rm out})\geq 10a$ & $2.73\%$\\
 3 & 30 & 30 & 1 & log-uniform & $a_{\rm out}(1-e_{\rm out})\geq 10a$ & $0.337\%$\\
 4 & 30 & 30 & 30 & log-uniform & $a_{\rm out}(1-e_{\rm out})\geq 5a$ & $1.40\%$\\
 5 & 30 & 30 & 30 & log-uniform & $a_{\rm out}(1-e_{\rm out})\geq 8a$ & $1.25\%$\\
 \hline 
\end{tabular}
\vspace*{0.3cm}
\caption{The settings of all the triple system runs. The inner semi-major axes $a$ range from 10 to 1000$\,$au. Note that Run 0 is the original one in \S\ref{subsec:triple}. Runs 1-5 are tests with $10^6$ triple systems each. All the other settings not mentioned in the table are the same as Run 0. Run 1 adopts a log-normal distribution for $a$, i.e. $\log_{10}(a)$ is assigned a mean of 1.7038 and a standard deviation of 1.52, inferred from fig.~13 in \cite{2010ApJS..190....1R}. Run 2 has unequal binary masses, which turns on the octupole-order secular effect and enhances the merger fraction. Run 3 uses a small tertiary mass, which reduces the merger fraction due to weaker LK effect. Run 4 and 5 adopt different cuts for the ratio between outer periastron and inner semi-major axis. Smaller ratio cuts (i.e. less hierarchical) lead to more systems with large tertiary perturbations, hence enhancing the merger fraction.}
\label{Tab:tests}
\end{table}

\bibliographystyle{aasjournal}
\bibliography{reference}



\end{document}